
\input phyzzx
\def\cL{{\cal L}}
\def\dplus{=\hskip-5pt \raise 0.7pt\hbox{${}_\vert$} ^{\phantom 7}}
\def\dplusup{=\hskip-5.1pt \raise 5.4pt\hbox{${}_\vert$} ^{\phantom 7}}
\def\dplus{=\hskip-4.8pt \raise 0.7pt\hbox{${}_\vert$} ^{\phantom 7}}

\def\pmb#1{\setbox0=\hbox{#1} \kern-.025em\copy0\kern-\wd0
\kern0.05em\copy0\kern-\wd0 \kern-.025em\raise.0433em\box0}

\def\cL{{\cal L}}
\def\cF{{\cal F}}
\def\cN{{\cal N}}
\def\cD{{\cal D}}
\def\cP{{\cal P}}
\def\cQ{{\cal Q}}

\font\mybb=msbm10 at 11pt

\def\bb#1{\hbox{\mybb#1}}

\def\bR {\bb{R}}

\def\bE {\bb{E}}

\def\bN {\bb{N}}

\def\bP {\bb{P}}

\def\bC {\bb{C}}

\hfuzz 1cm

\REF\gib{ G. W. Gibbons and P. J. Ruback, {\it  The Motion of extreme
Reissner-Nordstr\"om black holes in the low velocity limit},  Phys.
Rev. Lett. {\bf{57}} (1986) 1492.}

\REF\fer{R. C. Ferrell and D. M. Eardley, {\it Slow motion scattering
and coalescence of maximally charged black holes}, Phys. Rev. Lett.
{\bf{59}} (1987)n1617.}

\REF\strom{J. Michelson and A. Strominger, {\it Superconformal
 Multi-Black Hole Quantum
Mechanics}, JHEP 9909 (1999) 005; hep-th/9908044.}

\REF\guta{J. Gutowski and G. Papadopoulos, {\it The Dynamics of
Very Special Black Holes}, Phys. Lett. {\bf{B472}} (2000) 45;
hep-th/9910022.}

\REF\gutb{J. Gutowski and G. Papadopoulos, {\it Moduli Spaces for Four-
and Five- Dimensional Black Holes}, Phys. Rev. {\bf{D62}} (2000) 064023;
hep-th/0002242.}

\REF\stromb{R. Britto-Pacumio, A. Strominger and A. Volovich,
{\it Two-Black-Hole Bound States}, JHEP 0005 (2000) 020; hep-th/0004017.}

\REF\stringa{A. Comtet and G. W. Gibbons, {\it Bogomolny
Bounds for  Cosmic Strings}, Nucl. Phys. {\bf{B299}}: 719 (1988).}

\REF\stringb{B. R. Greene, A Shapere, C. Vafa and S-T Yau, {\it Stringy
Cosmic Strings and Noncompact Calabi-Yau Manifolds}, Nucl. Phys.
 {\bf{B337}}:1 (1990).}

\REF\papa{J. Gutowski and G. Papadopoulos, {\it Moduli Spaces and
Brane Solitons for M-Theory Compactifications on Holonomy $G_2$
Manifolds}; hep-th/0104105.}

\REF\keh{A. Kehagias, {\it N=2 Heterotic Stringy Cosmic Strings};
hep-th/9611110.}

\REF\ward{R. S. Ward, {\it Slowly Moving Lumps in the $\bC \bP^1$
Model in (2+1) dimensions}, Phys. Lett. {\bf{B158}} (1985) 424.}

\REF\spa{J. M. Speight, {\it Lump Dynamics in the  $\bC \bP^1$
model on the torus}, Commun. Math. Phys. {\bf{194}} (1998) 513;
 hep-th/9707101.}

\REF\spb{J. M. Speight, {\it Low Energy Dynamics of a  $\bC \bP^1$
lump on the sphere}, Phys. Rev. {\bf{D49}} (1994) 6914; hep-th/9712089.}

\REF\spc{J. M. Speight and I. A. B. Strachan, {\it Gravity Thaws
 the frozen Moduli of the  $\bC \bP^1$
lump}, Phys. Lett. {\bf{B457}} (1999) 12; hep-th/9903264.}

\REF\witbag{E. Witten and J. Bagger, {\it Quantization of Newton's
Constant in Certain Supergravity Theories}, Phys. Lett. {\bf{B115}}
 (1982) 202.}


\REF\stromd{A. Maloney, M. Spradlin and A. Strominger, {\it
Superconformal Multi-Black Hole Moduli Spaces in Four Dimensions};
hep-th/9911001.}

\REF\speightd{J. M. Speight, {\it The $L^2$ Geometry of spaces of
harmonic maps $S^2 \rightarrow S^2$
and $\bR \bP^2 \rightarrow \bR \bP^2$}; math.DG/0102038.}

\Pubnum{ \vbox{ \hbox{}\hbox{} } }
\pubtype{}
\date{September, 2001}
\titlepage
\title{ The Moduli Space of $\bC \bP^1$ Stringy Cosmic Lumps}
\author{J. Gutowski}
\address{Department of Physics,\break Queen Mary College,
\break Mile End,\break London E1 4NS}

\abstract {We examine the low-energy dynamics of $\bC \bP^1$ lumps
 coupled to gravity,
 taking into account the gravitational back-reaction of the
 spacetime geometry. We show that the
single lump moduli space is equipped with a three-dimensional metric,
and we derive stability bounds on the scalar coupling constant. We also
derive an expression for the multi-lump moduli space metric.}

\vskip 1.0 cm
\endpage
\pagenumber=2
\font\mybb=msbm10 at 12pt
\def\bb#1{\hbox{\mybb#1}}

\def\C{\mkern1mu\raise2.2pt\hbox{$\scriptscriptstyle|$}\mkern-7mu{\rm C}}

\def\l {\lambda}
\def\a{\alpha}
\def\pd{\partial_}
\def\pu{\partial^}
\def\b{\beta}
\def\g{\gamma}

\def\d{\delta}

\def\f{\phi}
\def\ps{\psi}
\def\t{\theta}
\def\O{\Omega}
\def\up{\Upsilon}
\def\bz{{\bar{z}}}
\def\bW{{\bar{W}}}
\def\bcP{{\bar{\cal P}}}
\def\bcQ{{\bar{\cal Q}}}
\def\x{\chi}
\def\bx{{\bar{\chi}}}
\def\bg{{\bar{\gamma}}}

\def\cN {{\cal {N}}}
\def\cT {{\cal {T}}}
\def\cV {{\cal {V}}}
\def\cD {{\cal {D}}}

\sequentialequations

\chapter{Introduction}

One particularly useful method of probing the low energy
 physics of objects arising in M-theory is the moduli space
approximation. This has been
 particularly apparent in the determining of the low energy
dynamics of black holes
of $\cN=2$ supergravity theories in four and five dimensions which are
associated with intersecting brane configurations in ten and
 eleven dimensions
compactified on Calabi-Yau manifolds
 [\gib , \fer, \strom, \guta, \gutb ].
Moreover, these geometries exhibit
a sufficiently high degree of symmetry that the associated sigma
 models typically admit supersymmetric extensions. On quantizing
these supersymmetric sigma models,
interesting results such as the existence of bound states in
a near horizon limit have been obtained [\stromb].

In addition to black holes, there are
other interesting supergravity solutions which are obtained from branes.
In particular, we shall concentrate on a class of
stringy cosmic string solutions
of four dimensional supergravity theories [\stringa , \stringb].
Such solutions may be obtained in various different ways. For example,
one may wrap an M5-brane on a supersymmetric co-associative 4-cycle
of a 7-manifold of $G_2$ holonomy. Compactifying along the $G_2$
 manifold, one obtains string solutions of $\cN=1$, $D=4$ supergravity
theory [\papa].
Alternatively,  stringy cosmic string solutions
of a four-dimensional $\cN=2$ theory may be obtained from
compactifying heterotic string theory on $K_3 \times T^2$ [\keh].
Compactifying further in the
direction parallel to the string,
one obtains lump solutions of a 3-dimensional theory.

In this paper we shall examine the low energy dynamics of lump solutions
of 3-dimensional gravity coupled to a complex scalar $\bC \bP^1$
sigma model obtained from the compactification of $\cN=2$, $D=4$
 supergravity, in which the scalar is a holomorphic rational function.
A considerable amount is known about the moduli space geometries of
such sigma models
in various fixed backgrounds [\ward, \spa, \spb]. One feature is
that some of the moduli are
{\it{frozen}} due to divergences in the integrals involved in
 the computations.
It has been argued in [\spc] that coupling the sigma model to gravity
provides a natural way of fixing an appropriate background geometry,
 and that
this geometry effectively un-freezes some of the moduli. A
 probe computation was used to derive a
six-dimensional metric on the moduli space of a single lump.
Although such computations  provide useful information
about the low energy
dynamics of objects in supergravity theories, they do
not generically take into account the back
reaction of the geometries to the motion of the bodies involved,
 and so many aspects of
the dynamics are neglected. This may be readily observed
 in the case of low energy black hole dynamics.

The plan of this paper is as follows. In section 2 we
describe some aspects of
the static solution. In section 3 we present
the moduli space perturbation
computation including the back-reaction of the 2+1 dimensional
spacetime, and show how the single lump moduli space
is described by a 3-dimensional
conical geometry. In section 4 we present some conclusions.

\chapter{The Static Solution}

The action which we shall consider is that of a
non-linear sigma model coupled to gravity in 2+1 dimensions.
It is given by

$$
S = \int \ d^3 x \  \sqrt{|g|} \big( R - 4 \l (1+|W|^2)^{-2}
 \pd{M}W \pu{M} \bW \big)
\eqn\act
$$
where $\l>0$ is a real constant, $g$ is the metric, and
 $W$ is a complex scalar.
The Einstein equations are given by
$$
G_{MN} +2 \l (1+|W|^2)^{-2} \big( \pd{S} W \pu{S} \bW g_{MN}
-2 \pd{(M}W \pd{N)}\bW \big) =0
\eqn\einst
$$
and the scalar equation obtained by varying $W$ is
$$
\pd{M} \big( \sqrt{|g|} (1+|W|^2)^{-2} \pu{M} \bW \big)
 +2 \sqrt{|g|} \bW (1+|W|^2)^{-3} \pd{M}W \pu{M} \bW =0 \ ,
\eqn\scala
$$
and taking the complex conjugate of this one obtains the
 $\bW$ scalar equation.

From now on, we shall adopt co-ordinates $x^M = \lbrack t,
 \ z, \ \bz \rbrack$, where $z$, $\bz$ are
complex co-ordinates, and $t$ is real. The solution which
corresponds to $N$ $\bC \bP^1$ lumps is given by
taking
$$
W = {\cP \over \cQ}
\eqn\statscal
$$
where $\cP=\cP(z)$ and $\cQ=\cQ(z)$ are degree $N$ polynomials
in $z$ with no common roots.
The spacetime metric is given by
$$
ds^2 = -dt^2 + \O^2 dz d\bz
\eqn\statmet
$$
where
$$
\O = (|\cP|^2+|\cQ|^2)^{-\l} \ .
\eqn\omexp
$$

\chapter{Moduli Metric Computation}

In order to perturb the $N$-lump system we allow the coefficients
 of $\cP$ and $\cQ$ to depend on $t$.
Such a solution automatically satisfies the scalar equations up
to first order in the velocities.
Hence, for a sigma model which is not coupled to gravity,
it is sufficient to simply substitute
this solution back into the action and obtain the terms
second order in the time derivatives, which
in turn defines the metric on the moduli space. However,
when one couples the sigma model to
gravity the Einstein equations must also be considered. It
is straightforward to see that
the $t z$ and $t \bz$ components are not satisfied to first
order. In order to rectify this it is
necessary to perturb the $t z$ and $t \bz$ components of
the metric according to
$$
\eqalign{
\d g_{tz} & = p_z
\cr
\d g_{t \bz} & = p_\bz \ ,}
\eqn\metpet
$$
where $p_z$ and $p_\bz$ are first order in the velocities
 and are to be determined by solving the Einstein
equations. Specifically, $p$ must satisfy
$$
\pd{z} \big[ \O^{-2} (\pd{z}p_\bz - \pd{\bz} p_z) +\l
(|\cP|^2+|\cQ|^2)^{-1}\big( \bcP \pd{t} \cP
-\cP \pd{t} \bcP + \bcQ \pd{t} \cQ - \cQ \pd{t} \bcQ \big)
\big] =0 \ ,
\eqn\constraint
$$
together with the complex conjugate of this expression.
This is solved by taking
$$
\O^{-2} (\pd{z}p_\bz - \pd{\bz} p_z) +\l (|\cP|^2+|\cQ|^2)^{-1}
\big( \bcP \pd{t} \cP
-\cP \pd{t} \bcP + \bcQ \pd{t} \cQ - \cQ \pd{t} \bcQ \big) = i \cL
\eqn\sola
$$
where $\cL$ is real, independent of $z$ and $\bz$ and is first
order in the velocities.
Defining
$$
\cD \equiv \cL (|\cP|^2+|\cQ|^2)^{-2\l} +i\l
(|\cP|^2+|\cQ|^2)^{-2\l-1}
\big( \bcP \pd{t} \cP
-\cP \pd{t} \bcP + \bcQ \pd{t} \cQ - \cQ \pd{t} \bcQ \big)\ ,
\eqn\hexp
$$
we note that \sola\ is solved by finding $\cV=\cV(z, \bz) \in \bC$
 such that
$$
\pd{z} \cV = {i \over 2} \cD \ ,
\eqn\vvexp
$$
and $p$ is given by
$$
\eqalign{
p_z & = {\bar{\cV}}+ \pd{z} \cT
\cr
p_\bz & = \cV + \pd{\bz} \cT\ ,}
\eqn\ppexp
$$
where $\cT=\cT(z, \bz) \in \bR$ is arbitrary.
It is then straightforward to compute the metric on the moduli
space. In particular,
substituting \metpet\ into \act\ one obtains the following
 second order terms

$$
\eqalign{
S_{(2)} & = \int dt \ dz \ d \bz \big[\cF+ \l
 (|\cP|^2+|\cQ|^2)^{-2\l-2} \big((|\cQ|^2-\l|\cP|^2)\pd{t}\cP \pd{t} \bcP
\cr
& +(|\cP|^2-\l |\cQ|^2)\pd{t} \cQ \pd{t} \bcQ
-(1+\l)(\cQ \bcP \pd{t} \bcQ \pd{t} \cP + \bcQ \cP \pd{t} \cQ \pd{t} \bcP)
 \big)
\cr
& - \O^{-2}  \big( \pd{z} p_\bz - \pd{\bz} p_z +\l
 (|\cP|^2+|\cQ|^2)^{-2\l -1}\big( \bcP \pd{t} \cP
-\cP \pd{t} \bcP + \bcQ \pd{t} \cQ - \cQ \pd{t} \bcQ \big) \big)^2
\ \big]\ ,}
\eqn\seca
$$
where $\cF$ integrates to give surface terms only;
$$
\eqalign{
\cF &= -4 \pd{z} \pd{\bz} ((|\cP|^2+|\cQ|^2)^{2 \l}p_z p_\bz )
+4 \pd{z} ((|\cP|^2+|\cQ|^2)^{2 \l}p_\bz \pd{z} p_\bz)
\cr
& +4 \pd{\bz} ((|\cP|^2+|\cQ|^2)^{2 \l} p_z \pd{\bz} p_z)
+4 \l \pd{z} ( (|\cP|^2+|\cQ|^2)^{-1} p_\bz (\bcP \pd{t} \cP +
\bcQ \pd{t} \cQ))
\cr
& +4 \l \pd{\bz} ( (|\cP|^2+|\cQ|^2)^{-1} p_z (\cP \pd{t} \bcP +
\cQ \pd{t} \bcQ))\ .}
\eqn\surf
$$
In particular, we shall only consider solutions in which the
contribution from $\cF$ vanishes.
Then substituting \sola\ into \seca\ we obtain the second order
 effective action
$$
\eqalign{
S_{(2)} & = \int dt \ dz \ d \bz \big[-4 \l^2
(|\cP|^2+|\cQ|^2)^{-2\l-1}(|\pd{t} \cP|^2+|\pd{t}\cQ|^2)
\cr
& +2\l (1+2\l) (|\cP|^2+|\cQ|^2)^{-2\l-2} |\cP \pd{t} \cQ
 - \cQ \pd{t} \cP|^2
+ (|\cP|^2+|\cQ|^2)^{-2\l} \cL^2 \big] \ .}
\eqn\secb
$$
Here $\cL$ is to be determined  by the requirement that $p$ is
 smooth, bounded and $\cF$
gives no contribution to the moduli space metric. It is
 possible to determine $p_z$
and $\cL$ explicitly in the simplest case of a single lump,
 where the polynomials
$\cP$ and $\cQ$ are of degree 1.

\section{Moduli Space of a $\bC \bP^1$ Lump}

It is straightforward to use the reasoning set out in the
previous section to compute the
metric on the moduli space of a single $\bC \bP^1$ lump.
In this case $\cP$ and $\cQ$ are degree 1 polynomials, and
 it is most convenient to write them as
$$
\eqalign{
\cP & = \sin {\t \over 2} e^{-{i \over 2}(\ps -\f)} (z-\g) +
 \b \cos {\t \over 2} e^{{i \over 2} (\ps +\f)}
\cr
\cQ & = \cos{\t \over 2} e^{-{i \over 2}(\ps +\f)} (z-\g) -
 \b \sin {\t \over 2} e^{{i \over 2}(\ps-\f)}\ ,}
\eqn\singlepol
$$
where $\g \in \bC$, $\b \in \bR^+$, and $0 \leq \t \leq \pi$,
$0 \leq \f \leq 2\pi$ and $0 \leq \ps <4 \pi$.
Then
$$
|\cP|^2+ |\cQ|^2 = \b^2 + |z-\g|^2 \ ,
\eqn\sauxa
$$
and
$$
\bcP \pd{t} \cP
-\cP \pd{t} \bcP + \bcQ \pd{t} \cQ - \cQ \pd{t} \bcQ = i \xi
 (\b^2 -|z-\g|^2) + \bx (z-\g) - \x (\bz-\bg)\ ,
\eqn\sauxb
$$
where
$$
\xi = {\dot{\ps}}+ \cos \t {\dot{\f}} \ ,
\eqn\xeq
$$
and
$$
\x = \b e^{i \ps} ({\dot{\t}}-i \sin \t {\dot{\f}})+{\dot{\g}}\ .
\eqn\xxeq
$$
Then \sola\ is solved by taking
$$
\eqalign{
\cV & = {i \over 2 (\bz - \bg)} (\b^2+ |z-\g|^2)^{-2\l} \big[\xi \b^2
 +{\cL + \xi \l \over 1-2\l}(\b^2+
|z-\g|^2) \big]
\cr
& -{\x \over 4} (\b^2+|z-\g|^2)^{-2\l}-{\bx \over 4(1-2\l) (\bz-\bg)^2}
(\b^2+|z-\g|^2)^{-2\l}
(\b^2+2\l|z-\g|^2)
\cr
\cT & = {1 \over 4(2\l-1)} (\b^2+|z-\g|^2)^{-2\l+1} \big({\x \over z-\g}
+{\bx \over \bz-\bg} \big)\ ,}
\eqn\solsingexp
$$
so that
$$
\eqalign{
p_z & = -{\bx \over 2} (\b^2+|z-\g|^2)^{-2\l}-{i \over 2 (z-\g)}
\big[\xi \b^2 (\b^2+|z-\g|^2)^{-2\l}
\cr
& \qquad \qquad +(1-2\l)^{-1} (\cL + \xi \l)(\b^2+|z-\g|^2)^{-2\l +1}
\big]
\cr
p_\bz & =  -{\x \over 2} (\b^2+|z-\g|^2)^{-2\l}+{i \over 2 (\bz-\bg)}
 \big[\xi \b^2 (\b^2+|z-\g|^2)^{-2\l}
\cr
& \qquad \qquad +(1-2\l)^{-1} (\cL + \xi \l)(\b^2+|z-\g|^2)^{-2\l +1}
 \big]\ .}
\eqn\pexp
$$
In order to ensure that $p$ remains bounded at $z=\g$ we set
$\cL = (\l-1)\xi$
so that
$$
\eqalign{
p_z & = {1 \over 2} (\b^2+|z-\g|^2)^{-2\l} \big( -\bx+
i \xi (\bz-\bg) \big)
\cr
p_\bz & = {1 \over 2} (\b^2+|z-\g|^2)^{-2\l} \big( -\x
-i \xi (z-\g) \big)\ .}
\eqn\pexpb
$$
We also impose
$\l>{1 \over 2}$ which ensures that the surface terms $\cF$
give no contribution to the moduli
space metric. In this case the volume of space takes the
finite value ${\pi \b^2 \over 2\l-1}$,
and the boundary at $|z|=\infty$ is a conical singularity
 of deficit angle $4\pi (1-\l)$.
So, on substituting this value of $\cL$ into \secb\ and
 performing the spatial integration we obtain
$$
S_{(2)} = \pi  \int \ dt \ (1-2\l)\b^{-4\l} {\dot{\b}}^2
+{(1-\l) \over 2(2\l-1)} \b^{2-4\l} ({\dot{\t}}^2
+\sin^2 \t {\dot{\f}}^2).
\eqn\secactsing
$$
Defining $r=\b^{1-2\l}$, we obtain the metric on the moduli
space (up to an overall constant conformal
factor of ${\pi \over 1-2\l}$) as
$$
ds^2 = dr^2 +{1 \over 2}(\l-1)r^2 (d\t^2+\sin^2 \t d \f^2)\ .
\eqn\modmetsing
$$
We observe that for $\l=3$ we have $ds^2 = ds^2(\bE^3)$, and for $\l=1$
the metric is degenerate.
For $\l \neq 1$, one may solve the geodesic equations
explicitly; in particular,
if $J^2 = r^4 ({\dot{\t}}^2+\sin^2 \t  {\dot{\f}}^2)$
 denotes the total conserved angular momentum,
then
$\b$ is given by
$$
\b = {1 \over \sqrt{\a^{-1} \big({1 \over 2}(\l-1)J^2
 +\a^2 (t+\d)^2 \big)^{{1 \over 2\l-1}}}}
\eqn\bexp
$$
where $\a \in \bR^+$ and $\d \in \bR$ are constants of
 integration. Hence, if ${1 \over 2}<\l<1$ then the moduli
space
approximation breaks down at some finite time, at which
 the spatial volume becomes infinite and this is a naked
curvature singularity. However,
if $\l>1$ then the moduli space approximation is
generically valid  for all time and
the volume of space is bounded above and tends to zero
 as $t \rightarrow \pm \infty$.  In particular,
we recall that \act\ may be obtained from a four-dimensional
action via an appropriate compactification, and it has been shown
in [\witbag] that in order for the scalar sigma model portion of
this action to describe a K\"ahler-Hodge manifold, one must have
$\l \in \bN$. So, for $\l \neq 1$, it follows that the moduli
metrics corresponding to single lump solutions of supersymmetric
theories have a conical geometry, except for $\l=3$ when $ds^2=
ds^2 (\bE^3)$.

\section{Multi-Lump Problem}

It is also possible to use the reasoning presented here to
 compute the metric on the moduli space
of multi-lump solutions. In fact, it is not even necessary
 to compute $p$ explicitly to determine
this. To see this we note that \sola\ may be written in
a form notation as
$$
dp = 2\cD dx \wedge dy
\eqn\stokes
$$
where $z=z+iy$, $\bz = x-iy$ and $x$, $y$ are real
 co-ordinates. Then for smooth $p$
such that $|p| \sim |z|^{-c}$ for $c >1$ as
$|z| \rightarrow \infty$, applying
Stokes' theorem to \stokes\ one obtains the constraint
$$
\int \cD(x,y)\  dx dy =0
\eqn\intconstr
$$
which fixes $\cL$ uniquely as
$$
\cL = -i \l {\big[\int \  dx dy \ (|\cP|^2+|\cQ|^2)^{-2\l-1}
\big( \bcP \pd{t} \cP
-\cP \pd{t} \bcP + \bcQ \pd{t} \cQ - \cQ \pd{t} \bcQ \big) \big]
\over \int \ dx dy \
 (|\cP|^2+|\cQ|^2)^{-2\l}} \ .
\eqn\curlyl
$$
We observe that the constraint  $\l > {1 \over 2N}$  is sufficient
to ensure the appropriate asymptotic
behaviour of $p$ as $|z| \rightarrow \infty$.
This constraint is satisfied if $\l \in \bN$,
 which corresponds to
supersymmetric theories which are of most interest.
We shall only consider such $\l$.
Even with this simplification, evaluating the integrals
 explicitly for generic $N \geq 2$
and $\l \in \bN$ is difficult. It is however instructive to
 examine the two-body problem.
In particular, suppose we consider the symmetric configuration
 given by
taking
$$
\eqalign{
\cP & = \sin {\t \over 2} e^{-{i \over 2}(\ps -\f)} H
 + \b \cos {\t \over 2} e^{{i \over 2} (\ps +\f)}
\cr
\cQ & = \cos{\t \over 2} e^{-{i \over 2}(\ps +\f)} H
 - \b \sin {\t \over 2} e^{{i \over 2}(\ps-\f)}\ ,}
\eqn\doublea
$$
where $\b \in \bR^+$, $0 \leq \t \leq \pi$, $0 \leq \f \leq 2\pi$
, $0 \leq \ps <4 \pi$
and
$$
H = z^2-\g
\eqn\twosym
$$
for $\g \in \bC$. The lumps are centred around $z = \pm \sqrt{\g}$.
Then
$$
|\cP|^2+ |\cQ|^2 = \b^2 + |H|^2 \ ,
\eqn\sauxaa
$$
and
$$
\bcP \pd{t} \cP
-\cP \pd{t} \bcP + \bcQ \pd{t} \cQ - \cQ \pd{t} \bcQ
= i \xi (\b^2 -|H|^2) + \bx H - \x {\bar{H}}\ ,
\eqn\sauxba
$$
where
$$
\xi = {\dot{\ps}}+ \cos \t {\dot{\f}} \ ,
\eqn\xeqa
$$
and
$$
\x = \b e^{i \ps} ({\dot{\t}}-i \sin \t {\dot{\f}})+{\dot{\g}}\ .
\eqn\xxeqa
$$
It is convenient to define for $c \geq 1$,
$$
\up_c = \int dx dy (\b^2+|H|^2)^{-c}\ .
\eqn\upsy
$$
With these conventions, $\cL$ is fixed as
$$
\cL = \l \xi \big( {2 \b^2 \up_{2\l+1} \over \up_{2\l}} -1 \big)\ ,
\eqn\topol
$$
and it follows that
$$
\eqalign{
S_{(2)} & = \l \int dt \big[ ({\dot{\theta}}^2+
\sin^2 \t {\dot{\f}}^2)({1 \over 2}\up_{2\l} -(1+2\l)\b^2 \up_{2\l+1}
+(1+2\l)\b^4 \up_{2\l+2})
\cr
&+2{\dot{\b}}^2 ( \up_{2\l+1}-(1+2\l)\b^2 \up_{2\l+2})
+({\dot{\g}} \bx +{\dot{\bar{\g}}} \x)(-2\l \up_{2\l+1}
+(1+2\l)\b^2\up_{2\l+2})
\cr
& +2\b^2 \xi^2(\up_{2\l+1}-(1+2\l)\b^2\up_{2\l+2}
+2 \l \b^2{\up_{2\l+1}{}^2 \over \up_{2\l}}) \big]\ .}
\eqn\compsec
$$
We observe that the double integral in \upsy\ may be rewritten
 for $c \in \bN$ as
$$
\up_c = {\pi^2 (2c-2)! \over 2^{2c-1} ((c-1)!)^2} \b^{1-2c}
 \ {}_2F_1 \big({1 \over 2}, c-{1 \over 2};1;
-{|\g|^2 \over \b^2} \big)\ ,
\eqn\hypergeom
$$
where ${}_2 F_1$ denotes the analytic continuation of the
 hypergeometric function defined
on $\bC / [1, \infty)$ with the branch cut along the positive real axis.

\chapter{Conclusions}

We have shown that the second order effective action which
 determines the low energy
dynamics of a single  $\bC \bP^1$ lump is simplified considerably
 when one includes
the back reaction of the spacetime geometry.
In particular, for stable solutions, we require that $\l>1$.
 In the case of
$\l = 3$ it is clear from the reasoning in
[\stromd] that the moduli space metric $ds^2 (\bE^3)$ admits
 a supersymmetric extension
to give a $\cN=4$ supersymmetric sigma model, even though
there are only three
moduli involved in \modmetsing\ . For $\l =3$, this result
 is consistent with
the correspondence between moduli metrics of supersymmetric
 gravity and
brane world-volume theories and supersymmetric sigma models.
For $\l \neq 3$, the moduli metric is
not of the form given in [\stromd], and the correspondence
is less clear.

 In addition, we have computed the metric on the moduli space
for a symmetric configuration of two lumps. The metric
 cannot be written in a simple closed form. It
would be interesting
to determine if there is a relationship between this metric
and the class of $SO(3) \times SO(3)$
invariant K\"ahler metrics
discussed in [\speightd] as it is clear that including the
 gravitational back-reaction alters
the form of the metrics. Furthermore, the methods used here
 may be easily
extended to probe the moduli dynamics of a symmetric configuration
 of $N$ lumps by making the
replacement
$z^2-\g \rightarrow z^N-\g$ in \twosym\ . It may also be possible
 to determine some of the
properties of more general non-symmetric
$N$-lump moduli space geometries for particular values of the
 coupling $\l$ from the
integral expressions presented here, using techniques similar
to those used to analyse
black hole moduli space geometries.

\vskip 1cm
\noindent{\bf Acknowledgements:} I thank
J. Gauntlett and especially G. Gibbons for useful conversations,
and also the organizers of the LMS Durham Symposia ``Special
 Structures in Differential Geometry''
during which part of this work was completed. J. G. is
 supported by an EPSRC
postdoctoral grant. This work is partially supported by
 SPG grant PPA/G/S/1998/00613.

\refout
\end